\documentclass[aps,pra,twocolumn,superscriptaddress,showpacs,floatfix]{revtex4-1}
\usepackage{lmodern,graphicx,amsmath,amssymb,amsfonts,multirow}
\usepackage[utf8]{inputenc}
\usepackage{gensymb}
\usepackage{xcolor}
\usepackage{longtable}

\newcommand\T{\rule{0pt}{2.6ex}}       
\newcommand\B{\rule[-1.2ex]{0pt}{0pt}} 

\begin{document}

\title{Optical bistability and nonlinear dynamics by saturation of cold Yb atoms in a cavity}

\author{Hannes Gothe}
\affiliation{Experimentalphysik, Universit\"at des Saarlandes, 66123 Saarbr\"ucken, Germany}
\affiliation{ICFO – The Institute of Photonic Sciences, Mediterranean Technology Park, Av. Carl Friedrich Gauss 3, 08860 Castelldefels (Barcelona), Spain}
\author{Tristan Valenzuela}
\affiliation{ICFO – The Institute of Photonic Sciences, Mediterranean Technology Park, Av. Carl Friedrich Gauss 3, 08860 Castelldefels (Barcelona), Spain}
\author{Matteo Cristiani}
\affiliation{ICFO – The Institute of Photonic Sciences, Mediterranean Technology Park, Av. Carl Friedrich Gauss 3, 08860 Castelldefels (Barcelona), Spain}
\author{Jürgen Eschner}
\affiliation{Experimentalphysik, Universit\"at des Saarlandes, 66123 Saarbr\"ucken, Germany}
\affiliation{ICFO – The Institute of Photonic Sciences, Mediterranean Technology Park, Av. Carl Friedrich Gauss 3, 08860 Castelldefels (Barcelona), Spain}

\date{\today}

\begin{abstract}

We observed optical bistability as well as oscillations in the upper bistable branch when cold ytterbium atoms are dispersively coupled to a high finesse optical cavity. Comparable previous observations were explained by atomic density oscillations in case of a Bose-Einstein-Condensate or optical pumping in case of a cold cloud of cesium. Both explanations do not apply in our case of thermal atoms with only a single ground state. We propose a simple two-level model including saturation to describe our experimental results. This paper introduces the experimental setup, derives the mentioned model and compares it to our experimental data. Good quantitative agreement supports our model.

\end{abstract}

\maketitle

\section{Introduction}
\label{sec:intro}

Bistability describes a hysteresis-like phenomenon where a system exposed to the same input values can reach two different states depending on its own history. In optics this typically occurs when a feedback-providing cavity is combined with a nonlinear medium. The non-linearity might be either absorptive as for saturable absorbers or dispersive as for media that change their refractive index with light intensity. Many different materials like atomic vapors, ruby, Kerr liquids and semiconductors were used and investigated starting in the 70's \cite{Szoeke1969,Gibbs1985}, mainly driven by the search for all-optical logic and switching devices. Nowadays, also quantum systems like ultra-cold atoms, Bose-Einstein-Condensates (BEC) and superconducting qubits \cite{Ritsch2013,Blais2004} are used as media while at the same time single photons became sufficient to produce bistable behavior \cite{Gupta2007,Ritter2009}, extending the scope of application from all-optical to quantum logic devices.

Here, we report the observation of optical bistability and oscillations in cold ytterbium extending similar studies with cold atoms\,\cite{Gupta2007,Lambrecht1995} and a BEC\,\cite{Brennecke2008,Ritter2009}. While the previous observations were explained by a competition of optical pumping and saturation\,\cite{Lambrecht1995} or atomic density oscillations\,\cite{Ritter2009}, these explanations do not apply in our case of a thermal cloud instead of a BEC and atoms with only a single ground state excluding optical pumping processes. Therefore, we present a simple two-level model incorporating saturation that describes our observations.

\begin{figure}[h]
\centering
\includegraphics[width=0.45\textwidth]{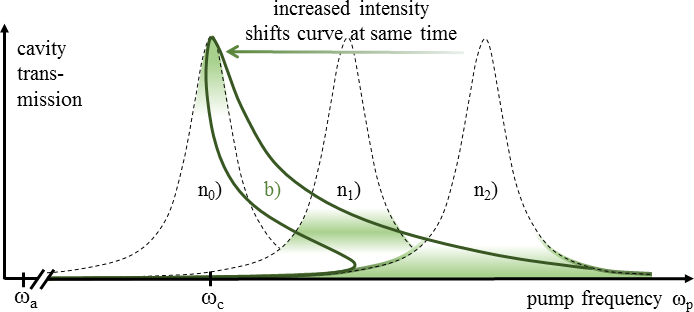}
\caption{
\label{fig:description}
The mechanism of dispersive bistability: An empty cavity shows a Lorentzian-shaped resonance curve (n$_0$) when a pump laser is scanned across the cavity resonance frequency ($\omega_\mathrm{c}$). Atoms inside the cavity act as a dispersive medium and shift the resonance curve depending on their number (n$_1$,n$_2$).
In turn, the effective atom number depends on the intra-cavity light intensity. With increasing intensity, more and more atoms get pumped to the excited state; they become transparent to the pump light and do no cause a dissipative shift any more.
This mutual influence between intra cavity light intensity and effective atom number alters the shape of the resonance curve.
When the cavity is filled with atoms and its resonance ($n_2$) is approached, the light intensity increases, bleaches a part of the atoms and therefore shifts the resonance frequency towards $\omega_\mathrm{c}$ (empty cavity). Each intensity level corresponds to a different position of the resonance curve as indicated by the green areas. For low intensities the resonance frequency shifts linearly with the intensity (n$_2$\,$\rightarrow$ n$_1$) causing the resulting resonance curve (b) to be tilted. For higher intensities ($I\gtrsim I_\mathrm{sat}$) the optical pumping saturates as most of the atoms are already bleached, and the resonance shift reduces (n$_1$\,$\rightarrow$ n$_0$) and the resulting curve (b) bends upwards at its tip.
}
\end{figure}

This paper starts by introducing our setup and measurement routine in section \ref{sec:exp}. Section \ref{sec:theory} derives a theoretical model, describes the principle working mechanism in figure \ref{fig:description}, and compares the model quantitatively to our measurements. Finally, section \ref{sec:oscillations} discusses the oscillations that occur in the upper bistable branch.

\section{Experimental setup}
\label{sec:exp}

\begin{figure}
\centering
\includegraphics[width=0.44\textwidth]{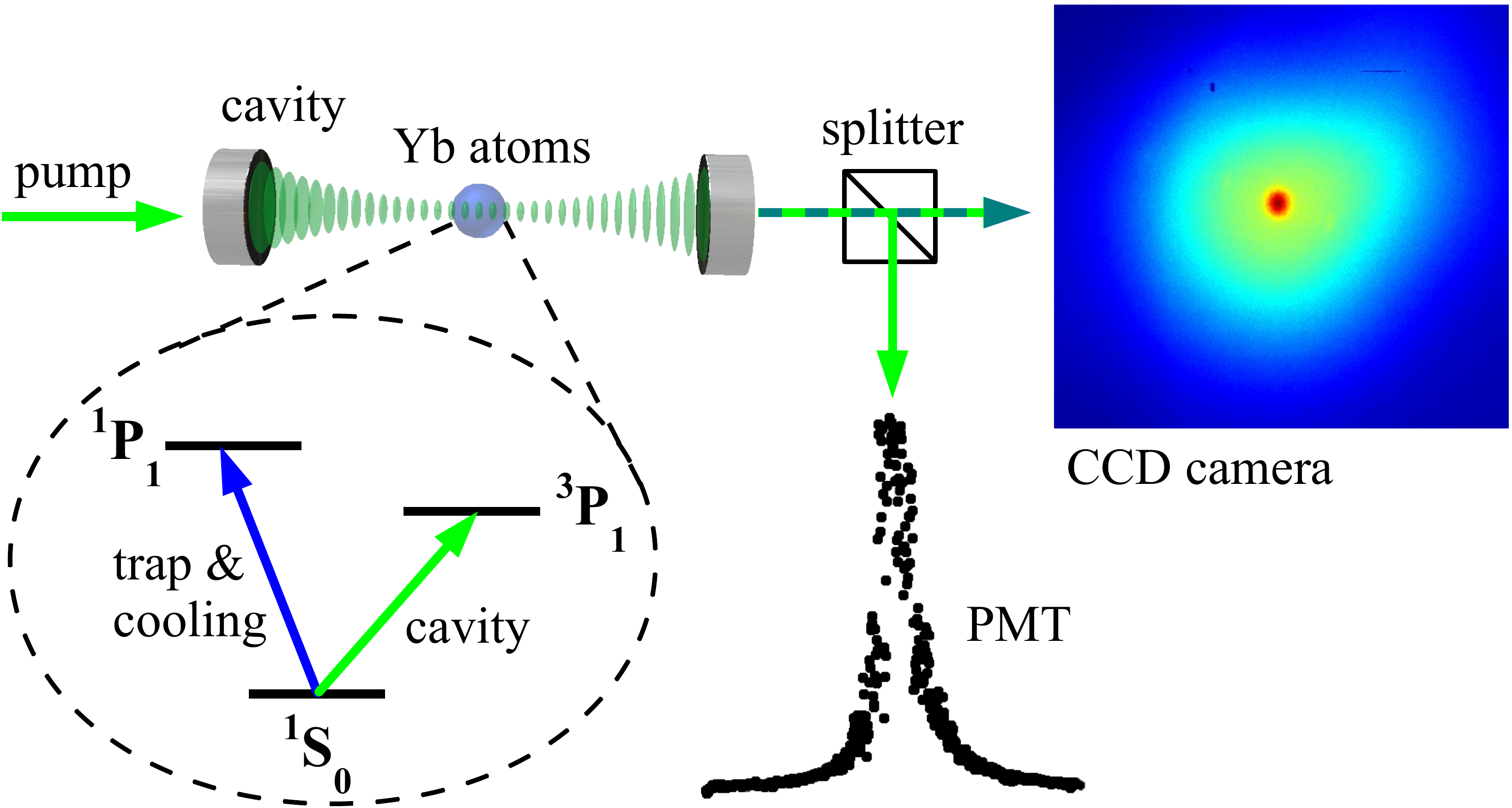}
\caption{
\label{fig:overview}
Setup: We pump a high-finesse optical cavity and scan over its resonance recording the transmission curve with a photo multiplier tube (PMT). We trap cold 	ytterbium atoms inside the resonator using the strong, blue $^1$S$_0{\to}^1$P$_1$ transition and tune the cavity field close to the narrower green $^1$S$_0{\to}^3$P$_1$ transition. This induces strong dispersive effects between atoms and cavity field and leads to a deformation of the transmission peak. With a camera we monitor the spacial overlap of the atoms (green) with the cavity mode (red dot).
}
\end{figure}

Our setup (figure \ref{fig:overview}) consists of a cloud of cold ytterbium ${}^{174}\mathrm{Yb}$ atoms that are continuously trapped inside a high finesse optical cavity. The magneto-optical trap operates on the ${}^1S_0{\rightarrow}{}^1P_1$ transition with an intensity of $\sim0.5\,I_\mathrm{sat}$, a detuning of $0.86\,\Gamma$, a MOT gradient of $36\,\mathrm{G/cm}$ and is constantly replenished by a Zeeman-slowed atomic beam. We trap about $10^7$ atoms at a temperature of 3\,mK\,\cite{Cristiani2010}.
The trap is permanently refilled from an atomic oven and the number of trapped atoms is adjusted by a mechanical shutter that controls the influx of new atoms.

\begin{table}
	\centering
	\begin{tabular}{lccccc}
																	&& wavelength 			&& linewidth						& saturation intensity \\
	Transition											&&  $\lambda$(nm)		&& $\Gamma/2\pi$ (MHz)	& $I_\mathrm{sat}$ (mW/cm$^2$)\B\\
	\hline
	 ${}^1S_0{\rightarrow}{}^1P_1$ 	&& $398.9$					&& $29$									& $59$\T\B\\
	 ${}^1S_0{\rightarrow}{}^3P_1$ 	&& $555.8$					&& $0.182$							& $0.14$\B\\
	\end{tabular}
\caption{\label{tab:transitions} Characteristics of the used ytterbium transitions.}
\end{table}

Our cavity consists of two highly reflective mirrors in Fabry-Perot configuration. It is resonant with the ${}^1S_0\rightarrow{}^3P_1$ transition and has a linewidth of $\kappa=2\pi\times70\,\mathrm{kHz}$ (finesse $\mathcal{F}=55\,000$) and a waist radius of $\mathrm{w}_0 = 90\,\mu$m.
We use a comparatively large cavity of 4.74\,cm which leaves enough space to trap atoms directly inside the cavity \footnote{Typical setups \cite{Purdy2010,Muenstermann1999a,Mabuchi1999} use short cavities of ${\simeq}100$\,$\mu$m in order to reach the strong coupling regime; the atoms are externally prepared and then moved into the cavity.}.

This long cavity leads to an atom-photon coupling of $g_0{}={}2\pi\times30\,\mathrm{kHz}$ corresponding to a cooperativity of $C=0.1$ which is below the strong-coupling threshold for single atoms. Nevertheless, for a sufficient number of atoms ($N\!>\!10$) we easily reach the collective strong coupling regime. 
We control and check the atom-cavity overlap by looking along the cavity axis using a CCD camera; the cavity mirrors are $90\,\%$ transparent for blue light. Therefore, we observe the cavity mode and the trap stray light at the same time (figure \ref{fig:overview}), and by centering the MOT ($\approx$1\,mm diameter) in the cavity we establish an overlap of $\lesssim$\,1\% between all trapped atoms and the cavity mode. Thus, a typical number of $N\simeq 10^5$ atoms is coupled to the cavity. 

\begin{figure}
\centering
\includegraphics[width=0.45\textwidth]{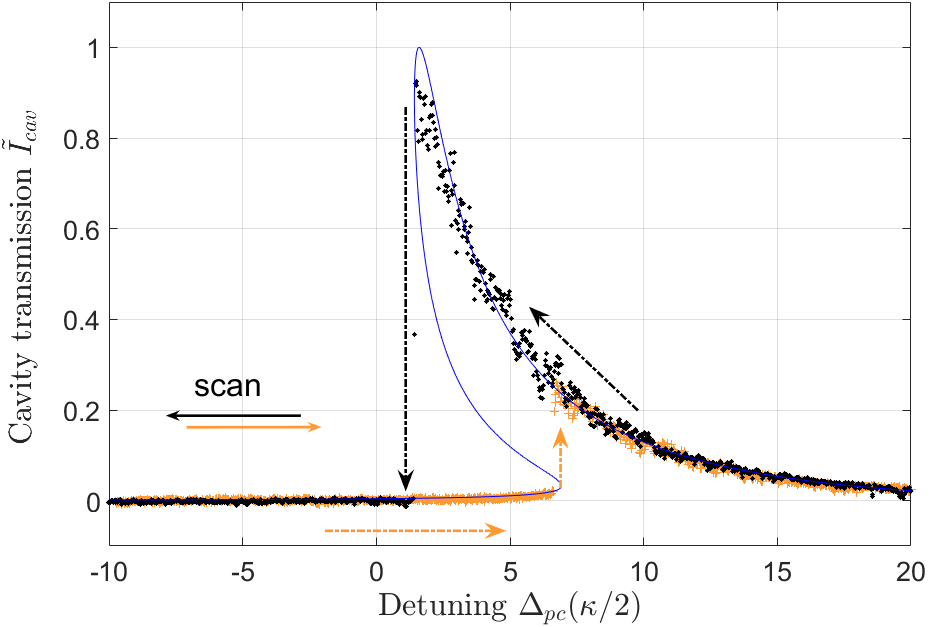}
\caption{
\label{fig:expdata}
The cavity transmission is measured in the presence of $\simeq$150,000 atoms by keeping the cavity fixed while changing the pump laser frequency. When scanning in positive direction we record the orange trace, whereas in negative direction we see the black one. We observe a bistable region ($\Delta_\mathrm{pc}=1 \ldots 7  \;\kappa/2$) where two transmission values are possible depending on the scan direction. The blue line depicts our model (\ref{eq:resonance_normalized}) with the parameters A=16 and S=9. 
}
\end{figure}

For the measurements presented in this paper we pump the fundamental mode of the cavity with a pump frequency $\omega_\mathrm{p}$ and a pump-cavity detuning $\Delta_\mathrm{pc}=\omega_\mathrm{p}-\omega_\mathrm{c}$ while we fix the cavity resonance $\omega_\mathrm{c}$ at a detuning of $\Delta_\mathrm{ca}=\omega_\mathrm{c}-\omega_\mathrm{a}$ = +30\,MHz (160\,$\Gamma$) from the atomic resonance $\omega_\mathrm{a}$ which makes the atoms a dispersive, rather than absorptive, medium inside the cavity. We then modulate the pump frequency in the vicinity of the cavity resonance scanning over 1234\,kHz during 0.2\,s at a pump power of 135\,$\mu\mathrm{W}$. We scan with increasing and decreasing pump frequency and record the transmission with a photo multiplier tube (PMT).  
For calibration each scan is performed twice: once with an empty cavity and a second time with atoms. The center of the Lorentzian peak in absence of atoms defines zero pump detuning. The presence of atoms inside the cavity alters the transmission profile significantly, showing typical signatures of optical bistability. For low atom numbers the Lorentzian curve becomes asymmetric and for higher atoms numbers we record steep rises and drops at different frequencies depending on the scan direction (figure \ref{fig:expdata}).

\section{Theoretical model}
\label{sec:theory}

In this section we develop a model that explains how the cavity transmission line shape is modified by the presence of atoms. Such a model has already been presented for a BEC and low photon numbers \cite{Ritter2009}. In our setup, however, we have a thermal cloud instead of  a BEC, and a mean intra-cavity photon number much higher than one. Therefore, we developed a different model which is inspired by the previously mentioned work, but which is formulated in a classical framework and adds saturation effects at higher intensities. Such a treatment was also suggested  by Lambrecht \textit{et al.} \cite{Lambrecht1995}.
The model describes the interplay of the intra-cavity light field and the atomic two-level population difference which is qualitively explained in figure \ref{fig:description}. Section \ref{sec:theory} discusses the steady-state solution of the model while Section \ref{sec:oscillations} describes the dynamics of the system appearing as oscillations around the steady state.

In an empty cavity that is pumped by a laser beam with frequency $\omega_\mathrm{p}$ and an electric field amplitude $E_{\mathrm{in}}$, the intra-cavity field $E_{\mathrm{cav}}$ evolves according to:
\begin{equation}
	\label{eq:cavityrate}
	\frac{\mathrm{d}}{\mathrm{d} t}E_{\mathrm{cav}}(t) = \mathrm{i} \frac{\kappa}{2} \sqrt{G} E_{\mathrm{in}} - \frac{\kappa}{2} E_{\mathrm{cav}}(t) + \mathrm{i} \Delta_\mathrm{pc} E_{\mathrm{cav}}(t),
\end{equation}
where $\kappa$ is the cavity linewidth (FWHM), $\Delta_\mathrm{pc} = \omega_\mathrm{p}-\omega_\mathrm{c}$ the detuning between the pump and the cavity frequency, $G = T / (1 - R)^{2} \simeq 360$ the power enhancement factor and $T$ and $R$ the transmittance and reflectivity of the cavity mirrors. The steady state solution of (\ref{eq:cavityrate}) is
\begin{equation}
	\label{eq:resonance_empty}
	\vert E_{\mathrm{cav}} \vert^2 = G \vert E_{\mathrm{in}} \vert^2 \frac{\kappa^2}{\kappa^2 + 4 \Delta_\mathrm{pc}^2}.
\end{equation}
This is a Lorentzian-shaped resonance curve, which is typical for an empty cavity.

Atoms inside the cavity resemble an optical medium and therefore increase the optical cavity length and shift the cavity resonance. This leads to an effective detuning
\begin{equation}
	\label{eq:Delta_n}
	\Delta_\mathrm{n} = \frac{g_0^2}{6 \Delta_\mathrm{ca}} \cdot {N_{\delta}}
\end{equation}
which adds to $\Delta_\mathrm{pc}$ in (\ref{eq:resonance_empty}) and depends on the population difference $N_{\delta}=N_S-N_P$ between ground and excited state and the shift due to a single maximally coupled atom $g_0^2/\Delta_\mathrm{ca}$ where $g_0$ is the atom cavity coupling rate and $\Delta_\mathrm{ca}=\omega_\mathrm{c}-\omega_\mathrm{a}$ the cavity atom detuning. A factor $\frac13$ arises from an average over the three possible directions of atomic orientation. Another factor $\frac12$ comes from an intensity average along the standing wave of the cavity.
If all atoms are in the ground state ($N_S=N=\,N_{\delta}$) they all contribute to the resonance shift. In contrast, if the transition is saturated ($N_{\delta}$=0) the atoms are transparent to this light. For two-level atoms $N_{\delta}$ is given by:
\begin{equation}
	N_{\delta}=N\left(1+\frac{\vert E_{\mathrm{cav}} \vert^2}{4\Delta_\mathrm{ca}^2 \Gamma^{-2} I_\mathrm{sat}}\right)^{-1}
\label{eq:N_diff}
\end{equation}
A factor $4\Delta_\mathrm{pa}^2 / \Gamma^{2}$ in front of the on-resonance saturation intensity $I_\mathrm{sat}$ accounts for the pump detuning with respect to the atomic resonance. Here we assume that $\Delta_\mathrm{pa}=\Delta_\mathrm{pc}+\Delta_\mathrm{ca}\simeq\Delta_\mathrm{ca}$ since the pump-cavity detuning $\Delta_\mathrm{pc}$ is small compared to $\Delta_\mathrm{pa}$ and $\Delta_\mathrm{ca}$.

Inserting (\ref{eq:N_diff}) into (\ref{eq:Delta_n}) and adding (\ref{eq:Delta_n}) to $\Delta_\mathrm{pc}$ in (\ref{eq:resonance_empty}) gives
\begin{equation*}
	\vert E_{\mathrm{cav}} \vert^2 = G \vert E_{\mathrm{in}} \vert^2 \frac{\kappa^2}{\kappa^2 + 4 \left[ \Delta_\mathrm{pc} + \frac{N g_0^2}{6 \Delta_\mathrm{ca}} \left(1+\frac{\vert E_{\mathrm{cav}} \vert^2}{4\Delta_\mathrm{ca}^2\Gamma^{-2} I_\mathrm{sat}}\right)^{-1} \right]^2}
\end{equation*}

In order to shorten this equation we introduce a normalized detuning $\tilde{\Delta}_\mathrm{pc}=2\Delta_\mathrm{pc}/\kappa$, an intra-cavity intensity $\tilde{I}_\mathrm{cav} = \vert E_{\mathrm{cav}} / E_{\mathrm{in}} \vert^2 / G$ that is normalized to its maximum value at resonance and two constants $A$ and $S$ leading to
\begin{equation}
	\label{eq:resonance_normalized}
	\tilde{I}_\mathrm{cav} = \frac{1}{1 + \left[ \tilde{\Delta}_\mathrm{pc} + A\left(1+S \tilde{I}_\mathrm{cav}\right)^{-1} \right]^2}
\end{equation}

The constant $A$ is given by $ N g_0^2 /( 3\Delta_\mathrm{ca} \kappa)$ and expresses the atom-cavity interaction strength. The saturation parameter $S=GI_\mathrm{in}\Gamma^2/(4\Delta_\mathrm{ca}^2I_\mathrm{sat})$ is the ratio between the maximum intra-cavity intensity at resonance and the detuned saturation intensity. In our experiments we set $A$ and $S$ via the atom number $N$ and the input intensity $I_\mathrm{in}$, respectively. The parameters of figure \ref{fig:expdata} ($N$\,$\simeq$\,150000, $I_\mathrm{in}$\,=\,135\,$\mu$W) correspond to $A$\,=\,20 and $S$\,=\,12. A fit results in $A$\,=\,16 and $S$\,=\,9. Taking into account the uncertainties especially for estimating the atom number from the MOT stray light we consider this a decent agreement.  

\begin{figure}
\centering
\includegraphics[width=0.41\textwidth]{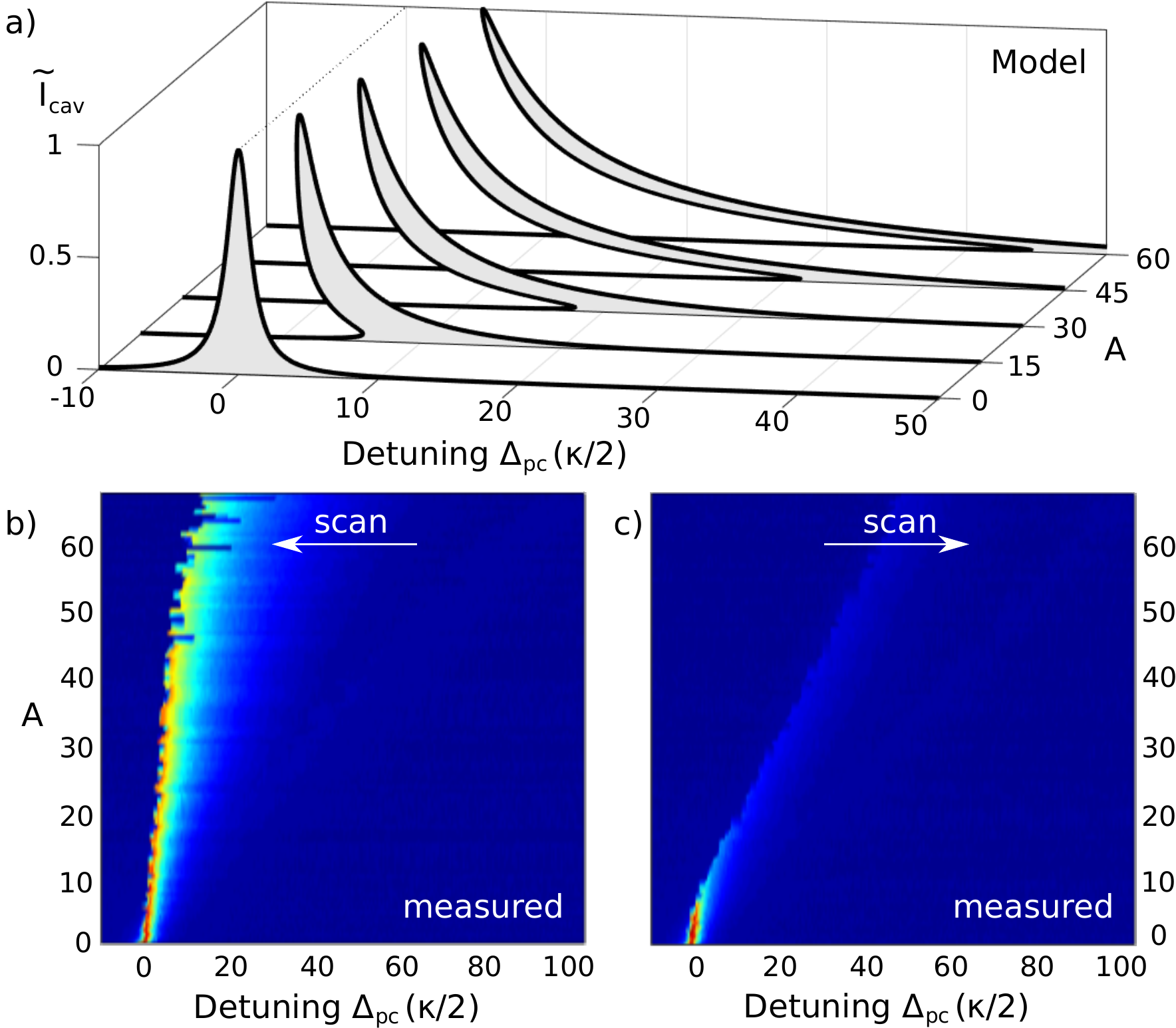}
\caption{
\label{fig:simulation}
Comparison of our model and the measured data for different atom numbers.
The upper plot a) shows calculations of the normalized cavity transmission $\tilde{I}_\mathrm{cav}$ according to (\ref{eq:resonance_normalized}) for different detunings $\tilde{\Delta}_\mathrm{pc}$ and atom-cavity interaction strength ($A$\,=\,0\dots60), where $A$ is proportional to the atom number. The parameter $S$ is set to 8. This yields a lorentzian curve for no atoms and increasingly tilted curves for larger atom numbers.
The lower plots show a corresponding set of experimental data with frequency scans across the cavity resonance in negative (b) and positive (c) scan direction. Each horizontal line represents one frequency scan at a fixed atom number (see figure \ref{fig:expdata}). The obtained transmission is color coded, blue representing low intensity and red high intensity. The atom number, indicated by parameter A, is varied along the vertical axis. 
}
\end{figure}

The parameters $A$ and $S$ define the shape of the resonance curve. The influence of $A$ is shown in figure \ref{fig:simulation}. For $A$=0 the resonance curve is symmetric and centered at zero detuning. For larger values the curve gets increasingly shifted and tilted. For sufficiently high atom numbers it exhibits a bistable region with multiple solutions. However, only one of these solutions is realized at a time when performing measurements in real systems. The solution that is closest to the previous state of the system is always preferred. This leads to a hysteresis effect where one branch of solutions is followed when entering the bistable region from one side but another branch when entering from the opposite side. In figure \ref{fig:simulation}b and \ref{fig:simulation}c the high or the low intensity branch is realized when scanning in negative or positive direction, respectively.

\section{Oscillations}
\label{sec:oscillations}

Many of our scans, though not all of them, show oscillations of the cavity transmission at a $\mu$s-timescale. The previous plots (figure \ref{fig:expdata}) were averaged to remove these fast oscillations. Now, we will focus on them.

\begin{figure}
\centering
\includegraphics[width=0.41\textwidth]{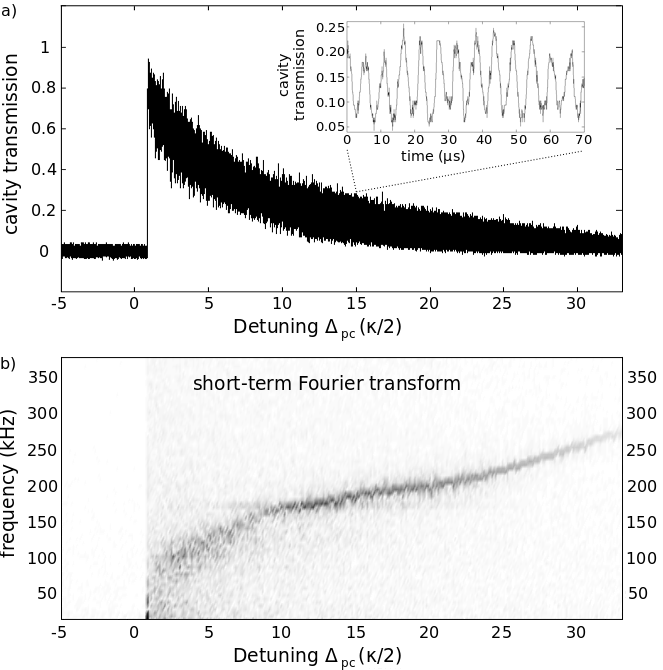}
\caption{
\label{fig:oscillations}
a) Oscillation of the cavity transmission during a scan with decreasing detuning. The scan was performed in 68\,ms with 250.000 atoms overlapped to the cavity mode. The inset shows a zoom at $\tilde{\Delta}_\mathrm{pc} = 15$ revealing oscillations at a $\mu$s-timescale. b) A short-term Fourier transform obtained from local sections of the transmission signal shows that the oscillation frequency is changing during the scan.    
}
\end{figure} 

The oscillations occur in both scanning directions throughout all the transmissive region. However, they are most evident in the upper bistable branch when scanning with decreasing detuning and using many atoms (250,000) as depicted in figure \ref{fig:oscillations}. Even higher atom numbers tend to collapse the upper branch as the system gets too unstable and abruptly settles to the lower branch. The frequency of the oscillation changes between 100 and 300\,kHz while scanning (figure \ref{fig:oscillations}b). When less atoms are used, the tail of the whole resonance shortens but the oscillation frequency at a certain detuning stays the same. Interestingly, the frequency is close to the atomic decay rate of 182\,kHz even though a change of the transmission properties should be limited by the cavity decay (70\,kHz). 

For comparison we numerically simulate a scan with the parameters of figure \ref{fig:oscillations}, using (\ref{eq:cavityrate}) (adding (\ref{eq:Delta_n}) to $\Delta_\mathrm{pc}$) for the evolution of the intracavity field and using
\begin{equation}
	\label{eq:population_rate}
	\frac{\mathrm{d}}{\mathrm{d} t}N_{\delta} = \frac{N}{\tau} - \frac{N_{\delta}}{\tau}\left(1+\frac{I_\mathrm{cav}}{I_\mathrm{sat}}\right)
\end{equation}
for the evolution of the population difference. These calculations also yield kHz-oscillations that change their frequency during the scan. However, the calculated frequencies are about four times smaller than the measured ones. A similar discrepancy was found in \cite{Ritter2009} and needs further investigation.

\section{Discussion}
\label{sec:discussion}

Our observations are almost identical to the ones previously found in a BEC\,\cite{Ritter2009} and cold cesium\,\cite{Lambrecht1995}, but they are obtained in a different physical system. Therefore, a different mechanism of the atom-light interplay must be at work. In our model the atom-cavity coupling diminishes when part of the atoms are pumped to the excited state. In the opto-mechanical model the coupling decreases when atoms move from an anti-node to a node of the cavity lattice and for cesium it increases when the atoms are pumped to higher Zeeman states. Nevertheless, all mechanisms lead to mathematically similar equations and similar phenomena.

Interestingly, our model also predicts oscillations but the calculated oscillation frequencies deviate from the observed ones by a factor of 4, similar to the result in \cite{Ritter2009}.

\bibliography{bistability}

\begin{thebibliography}{13}%
\makeatletter
\providecommand \@ifxundefined [1]{%
 \@ifx{#1\undefined}
}%
\providecommand \@ifnum [1]{%
 \ifnum #1\expandafter \@firstoftwo
 \else \expandafter \@secondoftwo
 \fi
}%
\providecommand \@ifx [1]{%
 \ifx #1\expandafter \@firstoftwo
 \else \expandafter \@secondoftwo
 \fi
}%
\providecommand \natexlab [1]{#1}%
\providecommand \enquote  [1]{``#1''}%
\providecommand \bibnamefont  [1]{#1}%
\providecommand \bibfnamefont [1]{#1}%
\providecommand \citenamefont [1]{#1}%
\providecommand \href@noop [0]{\@secondoftwo}%
\providecommand \href [0]{\begingroup \@sanitize@url \@href}%
\providecommand \@href[1]{\@@startlink{#1}\@@href}%
\providecommand \@@href[1]{\endgroup#1\@@endlink}%
\providecommand \@sanitize@url [0]{\catcode `\\12\catcode `\$12\catcode
  `\&12\catcode `\#12\catcode `\^12\catcode `\_12\catcode `\%12\relax}%
\providecommand \@@startlink[1]{}%
\providecommand \@@endlink[0]{}%
\providecommand \url  [0]{\begingroup\@sanitize@url \@url }%
\providecommand \@url [1]{\endgroup\@href {#1}{\urlprefix }}%
\providecommand \urlprefix  [0]{URL }%
\providecommand \Eprint [0]{\href }%
\providecommand \doibase [0]{http://dx.doi.org/}%
\providecommand \selectlanguage [0]{\@gobble}%
\providecommand \bibinfo  [0]{\@secondoftwo}%
\providecommand \bibfield  [0]{\@secondoftwo}%
\providecommand \translation [1]{[#1]}%
\providecommand \BibitemOpen [0]{}%
\providecommand \bibitemStop [0]{}%
\providecommand \bibitemNoStop [0]{.\EOS\space}%
\providecommand \EOS [0]{\spacefactor3000\relax}%
\providecommand \BibitemShut  [1]{\csname bibitem#1\endcsname}%
\let\auto@bib@innerbib\@empty
\bibitem [{\citenamefont {Szöke}\ \emph {et~al.}(1969)\citenamefont {Szöke},
  \citenamefont {Daneu}, \citenamefont {Goldhar},\ and\ \citenamefont
  {Kurnit}}]{Szoeke1969}%
  \BibitemOpen
  \bibfield  {author} {\bibinfo {author} {\bibfnamefont {A.}~\bibnamefont
  {Szöke}}, \bibinfo {author} {\bibfnamefont {V.}~\bibnamefont {Daneu}},
  \bibinfo {author} {\bibfnamefont {J.}~\bibnamefont {Goldhar}}, \ and\
  \bibinfo {author} {\bibfnamefont {N.~A.}\ \bibnamefont {Kurnit}},\ }\href
  {\doibase http://dx.doi.org/10.1063/1.1652866} {\bibfield  {journal}
  {\bibinfo  {journal} {Applied Physics Letters}\ }\textbf {\bibinfo {volume}
  {15}},\ \bibinfo {pages} {376} (\bibinfo {year} {1969})}\BibitemShut
  {NoStop}%
\bibitem [{\citenamefont {Gibbs}(1985)}]{Gibbs1985}%
  \BibitemOpen
  \bibfield  {author} {\bibinfo {author} {\bibfnamefont {H.~M.}\ \bibnamefont
  {Gibbs}},\ }\href@noop {} {\emph {\bibinfo {title} {Optical Bistability:
  Controlling Light with Light}}}\ (\bibinfo  {publisher} {Academic Press
  Inc},\ \bibinfo {year} {1985})\BibitemShut {NoStop}%
\bibitem [{\citenamefont {Ritsch}\ \emph {et~al.}(2013)\citenamefont {Ritsch},
  \citenamefont {Domokos}, \citenamefont {Brennecke},\ and\ \citenamefont
  {Esslinger}}]{Ritsch2013}%
  \BibitemOpen
  \bibfield  {author} {\bibinfo {author} {\bibfnamefont {H.}~\bibnamefont
  {Ritsch}}, \bibinfo {author} {\bibfnamefont {P.}~\bibnamefont {Domokos}},
  \bibinfo {author} {\bibfnamefont {F.}~\bibnamefont {Brennecke}}, \ and\
  \bibinfo {author} {\bibfnamefont {T.}~\bibnamefont {Esslinger}},\ }\href
  {\doibase 10.1103/RevModPhys.85.553} {\bibfield  {journal} {\bibinfo
  {journal} {Rev. Mod. Phys.}\ }\textbf {\bibinfo {volume} {85}},\ \bibinfo
  {pages} {553} (\bibinfo {year} {2013})}\BibitemShut {NoStop}%
\bibitem [{\citenamefont {Blais}\ \emph {et~al.}(2004)\citenamefont {Blais},
  \citenamefont {Huang}, \citenamefont {Wallraff}, \citenamefont {Girvin},\
  and\ \citenamefont {Schoelkopf}}]{Blais2004}%
  \BibitemOpen
  \bibfield  {author} {\bibinfo {author} {\bibfnamefont {A.}~\bibnamefont
  {Blais}}, \bibinfo {author} {\bibfnamefont {R.-S.}\ \bibnamefont {Huang}},
  \bibinfo {author} {\bibfnamefont {A.}~\bibnamefont {Wallraff}}, \bibinfo
  {author} {\bibfnamefont {S.~M.}\ \bibnamefont {Girvin}}, \ and\ \bibinfo
  {author} {\bibfnamefont {R.~J.}\ \bibnamefont {Schoelkopf}},\ }\href
  {\doibase 10.1103/PhysRevA.69.062320} {\bibfield  {journal} {\bibinfo
  {journal} {Phys. Rev. A}\ }\textbf {\bibinfo {volume} {69}},\ \bibinfo
  {pages} {062320} (\bibinfo {year} {2004})}\BibitemShut {NoStop}%
\bibitem [{\citenamefont {Gupta}\ \emph {et~al.}(2007)\citenamefont {Gupta},
  \citenamefont {Moore}, \citenamefont {Murch},\ and\ \citenamefont
  {Stamper-Kurn}}]{Gupta2007}%
  \BibitemOpen
  \bibfield  {author} {\bibinfo {author} {\bibfnamefont {S.}~\bibnamefont
  {Gupta}}, \bibinfo {author} {\bibfnamefont {K.~L.}\ \bibnamefont {Moore}},
  \bibinfo {author} {\bibfnamefont {K.~W.}\ \bibnamefont {Murch}}, \ and\
  \bibinfo {author} {\bibfnamefont {D.~M.}\ \bibnamefont {Stamper-Kurn}},\
  }\href {\doibase 10.1103/PhysRevLett.99.213601} {\bibfield  {journal}
  {\bibinfo  {journal} {Phys. Rev. Lett.}\ }\textbf {\bibinfo {volume} {99}},\
  \bibinfo {pages} {213601} (\bibinfo {year} {2007})}\BibitemShut {NoStop}%
\bibitem [{\citenamefont {Ritter}\ \emph {et~al.}(2009)\citenamefont {Ritter},
  \citenamefont {Brennecke}, \citenamefont {Baumann}, \citenamefont {Donner},
  \citenamefont {Guerlin},\ and\ \citenamefont {Esslinger}}]{Ritter2009}%
  \BibitemOpen
  \bibfield  {author} {\bibinfo {author} {\bibfnamefont {S.}~\bibnamefont
  {Ritter}}, \bibinfo {author} {\bibfnamefont {F.}~\bibnamefont {Brennecke}},
  \bibinfo {author} {\bibfnamefont {K.}~\bibnamefont {Baumann}}, \bibinfo
  {author} {\bibfnamefont {T.}~\bibnamefont {Donner}}, \bibinfo {author}
  {\bibfnamefont {C.}~\bibnamefont {Guerlin}}, \ and\ \bibinfo {author}
  {\bibfnamefont {T.}~\bibnamefont {Esslinger}},\ }\href {\doibase
  10.1007/s00340-009-3436-9} {\bibfield  {journal} {\bibinfo  {journal} {Appl.
  Phys. B}\ }\textbf {\bibinfo {volume} {95}},\ \bibinfo {pages} {213}
  (\bibinfo {year} {2009})}\BibitemShut {NoStop}%
\bibitem [{\citenamefont {Lambrecht}\ \emph {et~al.}(1995)\citenamefont
  {Lambrecht}, \citenamefont {Giacobino},\ and\ \citenamefont
  {Courty}}]{Lambrecht1995}%
  \BibitemOpen
  \bibfield  {author} {\bibinfo {author} {\bibfnamefont {A.}~\bibnamefont
  {Lambrecht}}, \bibinfo {author} {\bibfnamefont {E.}~\bibnamefont
  {Giacobino}}, \ and\ \bibinfo {author} {\bibfnamefont {J.}~\bibnamefont
  {Courty}},\ }\href {\doibase https://doi.org/10.1016/0030-4018(94)00493-E}
  {\bibfield  {journal} {\bibinfo  {journal} {Optics Communications}\ }\textbf
  {\bibinfo {volume} {115}},\ \bibinfo {pages} {199 } (\bibinfo {year}
  {1995})}\BibitemShut {NoStop}%
\bibitem [{\citenamefont {Brennecke}\ \emph {et~al.}(2008)\citenamefont
  {Brennecke}, \citenamefont {Ritter}, \citenamefont {Donner},\ and\
  \citenamefont {Esslinger}}]{Brennecke2008}%
  \BibitemOpen
  \bibfield  {author} {\bibinfo {author} {\bibfnamefont {F.}~\bibnamefont
  {Brennecke}}, \bibinfo {author} {\bibfnamefont {S.}~\bibnamefont {Ritter}},
  \bibinfo {author} {\bibfnamefont {T.}~\bibnamefont {Donner}}, \ and\ \bibinfo
  {author} {\bibfnamefont {T.}~\bibnamefont {Esslinger}},\ }\href {\doibase
  10.1126/science.1163218} {\bibfield  {journal} {\bibinfo  {journal}
  {Science}\ }\textbf {\bibinfo {volume} {322}},\ \bibinfo {pages} {235}
  (\bibinfo {year} {2008})}\BibitemShut {NoStop}%
\bibitem [{\citenamefont {Cristiani}\ \emph {et~al.}(2010)\citenamefont
  {Cristiani}, \citenamefont {Valenzuela}, \citenamefont {Gothe},\ and\
  \citenamefont {Eschner}}]{Cristiani2010}%
  \BibitemOpen
  \bibfield  {author} {\bibinfo {author} {\bibfnamefont {M.}~\bibnamefont
  {Cristiani}}, \bibinfo {author} {\bibfnamefont {T.}~\bibnamefont
  {Valenzuela}}, \bibinfo {author} {\bibfnamefont {H.}~\bibnamefont {Gothe}}, \
  and\ \bibinfo {author} {\bibfnamefont {J.}~\bibnamefont {Eschner}},\ }\href
  {\doibase 10.1103/PhysRevA.81.063416} {\bibfield  {journal} {\bibinfo
  {journal} {Phys. Rev. A}\ }\textbf {\bibinfo {volume} {81}},\ \bibinfo
  {pages} {063416} (\bibinfo {year} {2010})}\BibitemShut {NoStop}%
\bibitem [{Note1()}]{Note1}%
  \BibitemOpen
  \bibinfo {note} {Typical setups \cite
  {Purdy2010,Muenstermann1999a,Mabuchi1999} use short cavities of ${\simeq
  }100$\protect \tmspace +\thinmuskip {.1667em}$\mu $m in order to reach the
  strong coupling regime; the atoms are externally prepared and then moved into
  the cavity.}\BibitemShut {Stop}%
\bibitem [{\citenamefont {Purdy}\ \emph {et~al.}(2010)\citenamefont {Purdy},
  \citenamefont {Brooks}, \citenamefont {Botter}, \citenamefont {Brahms},
  \citenamefont {Ma},\ and\ \citenamefont {Stamper-Kurn}}]{Purdy2010}%
  \BibitemOpen
  \bibfield  {author} {\bibinfo {author} {\bibfnamefont {T.~P.}\ \bibnamefont
  {Purdy}}, \bibinfo {author} {\bibfnamefont {D.~W.~C.}\ \bibnamefont
  {Brooks}}, \bibinfo {author} {\bibfnamefont {T.}~\bibnamefont {Botter}},
  \bibinfo {author} {\bibfnamefont {N.}~\bibnamefont {Brahms}}, \bibinfo
  {author} {\bibfnamefont {Z.-Y.}\ \bibnamefont {Ma}}, \ and\ \bibinfo {author}
  {\bibfnamefont {D.~M.}\ \bibnamefont {Stamper-Kurn}},\ }\href {\doibase
  10.1103/PhysRevLett.105.133602} {\bibfield  {journal} {\bibinfo  {journal}
  {Phys. Rev. Lett.}\ }\textbf {\bibinfo {volume} {105}},\ \bibinfo {pages}
  {133602} (\bibinfo {year} {2010})}\BibitemShut {NoStop}%
\bibitem [{\citenamefont {Münstermann}\ \emph {et~al.}(1999)\citenamefont
  {Münstermann}, \citenamefont {Fischer}, \citenamefont {Pinkse},\ and\
  \citenamefont {Rempe}}]{Muenstermann1999a}%
  \BibitemOpen
  \bibfield  {author} {\bibinfo {author} {\bibfnamefont {P.}~\bibnamefont
  {Münstermann}}, \bibinfo {author} {\bibfnamefont {T.}~\bibnamefont
  {Fischer}}, \bibinfo {author} {\bibfnamefont {P.~W.~H.}\ \bibnamefont
  {Pinkse}}, \ and\ \bibinfo {author} {\bibfnamefont {G.}~\bibnamefont
  {Rempe}},\ }\href {\doibase 10.1016/S0030-4018(98)00596-3} {\bibfield
  {journal} {\bibinfo  {journal} {Opt. Commun.}\ }\textbf {\bibinfo {volume}
  {159}},\ \bibinfo {pages} {63} (\bibinfo {year} {1999})}\BibitemShut
  {NoStop}%
\bibitem [{\citenamefont {Mabuchi}\ \emph {et~al.}(1999)\citenamefont
  {Mabuchi}, \citenamefont {Ye},\ and\ \citenamefont {Kimble}}]{Mabuchi1999}%
  \BibitemOpen
  \bibfield  {author} {\bibinfo {author} {\bibfnamefont {H.}~\bibnamefont
  {Mabuchi}}, \bibinfo {author} {\bibfnamefont {J.}~\bibnamefont {Ye}}, \ and\
  \bibinfo {author} {\bibfnamefont {H.~J.}\ \bibnamefont {Kimble}},\ }\href
  {\doibase 10.1007/s003400050751} {\bibfield  {journal} {\bibinfo  {journal}
  {Appl. Phys. B}\ }\textbf {\bibinfo {volume} {68}},\ \bibinfo {pages} {1095}
  (\bibinfo {year} {1999})}\BibitemShut {NoStop}%
\end{thebibliography}%

\end{document}